\documentclass[preprint,showpacs,preprintnumbers,amsmath,amssymb,prb]{revtex4-1}
\usepackage{graphicx}
\usepackage{dcolumn}
\usepackage{bm}
\usepackage{epsfig}
\usepackage{subfigure}
\begin{document}

\title{A water-like model under confinement for hydrophobic and hydrophilic  
particle-plate interaction 
potentials}

\author{Leandro B. Krott}
\address{Instituto de F\'{i}sica, Universidade Federal do Rio 
Grande do Sul, 91501-970, Porto Alegre, Rio Grande do Sul}

\author{Marcia C. Barbosa}
\address{Instituto de F\'{i}sica, Universidade Federal do Rio 
Grande do Sul, 91501-970, Porto Alegre, Rio Grande do Sul}

\date{\today}

\begin{abstract}

Molecular dynamic simulations were employed to study a water-like model 
confined between hydrophobic and hydrophilic plates. The phase behavior of 
this system is obtained for different distances between the plates and particle-plate
potentials. For both hydrophobic and hydrophilic walls there are the formation of
layers. Crystallization occurs at lower temperature at the contact layer than at the 
middle layer. In addition, the melting temperature decreases as the plates become more
hydrophobic. Similarly, the temperatures of maximum density and extremum diffusivity decrease
with hydrophobicity.

\end{abstract}
\maketitle

\section{\label{sec1}Introduction}

Bulk water presents a peculiar complexity on its properties.
While in most materials the decrease of the temperature results in a monotonic increase 
of the density, the water, at ambient pressure, has a maximum in its density
at $4^oC$~\cite{Wa64,Ke75,An76}. Furthermore, for usual liquids the response functions increase with
the increase of temperature, while water exhibits an anomalous increase of compressibility
~\cite{Sp76,Ka79} 
between $0.1$ MPa and $190$ MPa and, at atmospheric pressure, an increase of isobaric
heat capacity upon cooling~\cite{An82,To99}. 
The anomalous behavior of water 
are not only related with thermodynamic functions, the 
diffusion coefficient for water 
has a maximum at $4^oC$ for $1.5\;atm$~\cite{An76,Pr87}, whereas for
normal liquids it increases with the decrease of pressure. The 
anomalies have been explained
in the framework of the existence of a liquid-liquid phase
transition ending in a second critical point. This critical 
point is, however, hidden in a region of the pressure-temperature 
phase diagram where homogeneous nucleation takes place,
and the two liquid phases do not equilibrate~\cite{Po92}. 

The difficulty of finding the liquid-liquid critical point has been circumvent
by experiments performed in confined 
systems. In these
systems the presence of criticality in
the bulk  has been associated with
a dynamic transition between liquids of 
different viscosities~\cite{Li05, Ch10b, Ch06, Ma08}. 

The study of water in confined geometries is, however,  important 
not only to understand its anomalous properties but also to learn about 
essential processes to the existence of life, like enzymatic 
activity of proteins~\cite{Ch06b, Fr08, Ku12}.
Confined water plays an important role in many other areas like 
chemistry, engineering
and geology. For these systems is very important to understand
the effect in the pressure-temperature phase
diagram of the size of the confinement and the hydrophobicity of the wall.

Experiments with water in confined geometries employing NMR~\cite{Ha95, Ov92} and X-ray 
diffraction~\cite{Mo99, Mo03b} show two complementary important
findings. First,  the pore size
has important influence on the freezing and melting temperature.
Next, the freezing in these systems is not uniform. Water
form layers inside the pores~\cite{Er11,De10,Ja08,Mo97}
that do not freeze at the same temperature~\cite{Be93} but the middle 
layers crystallize 
before the wall layers~\cite{Er11,Mo97}.

Less clear than the pore size effect is
the water-wall interaction effect on 
the melting temperature. Experimental studies show contradictory
results. While Akcakayiran et al.~\cite{Ak08} using calorimetry studies
of water in pores with phosphonic, sulfonic and carboxylic acids show
that the melting temperature is not affected by the change of surface, Deschamps et
 al.~\cite{De10} and Jelassi et al.~\cite{Je11} show that for water
confined in hydrophobic nanopores the liquid states 
persists to temperatures lower than in bulk and in 
hydrophilic confinement.
These observations are confirmed by  
X-ray and neutron diffraction~\cite{Je11, Ba08}.

Simulations agree with the
experiments in two points. First, the melting temperature of confined water
decreases as the system becomes more restrict by decreasing
the pore size or distance between plates~\cite{Gi09a,Mo12}.
Second, the system forms layers~\cite{Za03b, Ha10, Bo12a}
where not all the confined water crystallizes~\cite{Gi09,Mo12}. The
crystallized layer along the wall is in contact with a pre-wetting liquid
layer and some systems present the formation 
of layers where water just crystallizes partially~\cite{Gi09,Ku05,Ga00,So11}.

Simulations also show controversial results for the effect
of hydrophobicity in the melting temperature. While
results for SPC/E water show that the 
melting temperature for hydrophobic plates is lower than
the bulk and higher than for hydrophilic walls, for 
mW model no difference between  the
melting temperature due to the hydrophobicity~\cite{Mo12} is found.

In addition to the melting line, other thermodynamic
properties have been explored by simulations. In confined systems  the TMD 
occurs at lower temperatures for hydrophobic confinement~\cite{Gi09, Ku05} and 
at higher temperatures for hydrophilic confinement~\cite{Ca09} when compared with the bulk. 
The diffusion coefficient, $D$,
in the direction parallel to the plates exhibit an anomalous
behavior as observed in bulk water. However the temperatures of 
the the maximum and minimum of $D$ are lower than in bulk water~\cite{Ku05}.
In the direction perpendicular
to the plates, no diffusion anomalous behavior is 
observed~\cite{Ha08}.

In this paper we propose that many of the properties described
above can be explained in the framework of the competition
between the particle-particle interaction and the particle-plate
interaction potentials. 
For that purpose, we  study a 
core-softened fluid~\cite{Ol06a,Ol06b}
confined between parallel plates \cite{Kr13}. 
The particle-plate interaction is then varied from a very hydrophobic to 
hydrophilic interaction. The pressure and temperature
location of the  anomalies, melting and layering 
are compared with the bulk system for the different particle-plate interactions.

The paper is organized as follows: in Sec. II we introduce the model; 
in Sec. III the methods and simulation details are described; the 
results are given in Sec. IV; and  conclusions are presented in Sec. V.

\section{\label{sec:model} The Model}

\begin{figure}[!htb]
  \begin{centering}
\includegraphics[clip=true,width=12cm]{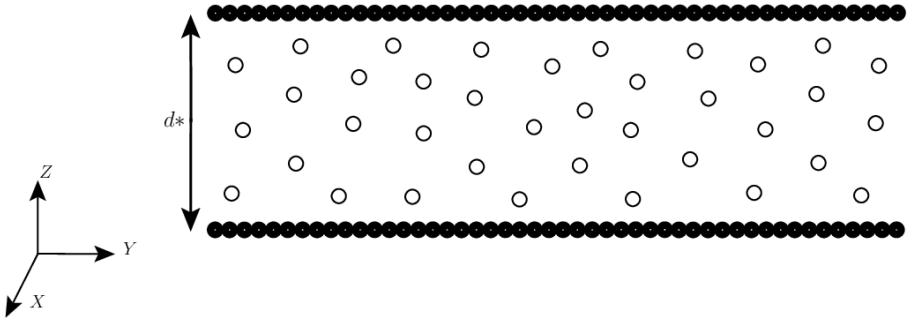}\par
  \end{centering}
  \caption{Schematic depiction of the particles confined between two plates
separated by a distance $d^*$.}
\label{model_figure}
\end{figure}

We study systems with $N$ particles of diameter $\sigma$ confined 
between two fixed plates. These plates are formed by particles of diameter
$\sigma$ organized in a square lattice of area $L^2$. The center-to-center plates 
distance is $d^* = d/\sigma$. A schematic depiction of the system is 
shown in Fig.~\ref{model_figure}. 

The particles of the fluid interact between them through an isotropic effective 
potential given by 
\begin{equation}\label{eq_potential}
 \centering
  \frac{U(r)}{\epsilon} = 4\left[ \left( \frac{\sigma}{r} \right)^{12} - 
\left( \frac{\sigma}{r} \right)^{6} \right] + 
a\exp\left[-\frac{1}{c^2}\left(\frac{r-r_0}{\sigma}\right)^2\right] \;\;.
\end{equation}

\begin{figure}[!htb]
  \begin{centering}
\includegraphics[clip=true,width=9cm]{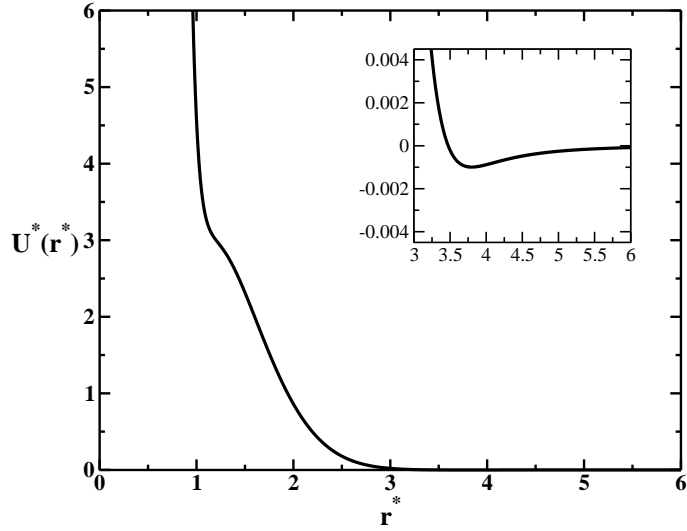}\par
    \par
  \end{centering}
  \caption{Isotropic effective potential (Eq.~\ref{eq_potential}) of 
interaction between the water-like particles. The energy and the 
distances are in dimensionless units, $U^* = U/\epsilon$ and 
$r^* = r/\sigma$ and the parameters are $a = 5$, $r_0/\sigma = 0.7$ 
and $c = 1$. The inset shows a zoom in the very small attractive part 
of the potential.}
  \label{potential_alan}
\end{figure}

The first term is a standard Lennard-Jones (LJ) $12-6$ potential 
with $\epsilon$ depth plus a Gaussian well centered on radius 
$r = r_0$ and width $c$. The parameters used are given by $a = 5$, $r_0/\sigma = 0.7$ 
and $c = 1$. 
This potential has two length scales with a repulsive shoulder at  
$r/\sigma \approx 1 $ and a very small attractive well at 
$r/\sigma \approx 3.8$ (Fig.~\ref{potential_alan}). This potential
represents in an effective way the tetramer-tetramer interaction
forming 
open and closed structures~\cite{URL}. The pressure versus temperature
phase diagram of this system in the bulk  was studied by 
Oliveira et~al.~\cite{Ol06a, Ol06b}. Density and diffusion anomalous behavior was found
for the bulk model.

In order to 
check the effect of hydrophobicity and confinement, five types of particle-plate 
interaction potentials were studied 
namely the repulsive of twenty-forth power (R24), of sixth power (R6), 
Weeks-Chandler-Andersen (WCA)~\cite{WCA_LJ}, 
a weak attractive (WAT) and a strong attractive (SAT).
The three first of them are purely repulsive potentials, while the other two 
have an attractive part. Our simulations are done in reduced units, where
$U^* = U/\epsilon$ and $r^* = r/\sigma$. Three purely repulsive potentials are used:
R6, R24 and WCA. The equation of the more repulsive potential (R6) is

\begin{equation}\label{eq_potentialr6}
U_{R6}^* = \frac{U_{R6}}{\epsilon} = \left\{ \begin{array}{ll}
 A_1\left(\sigma/r \right)^{6} + A_2\left(r/\sigma\right) -
\varepsilon_{1}  , \qquad r \le r_{c1} \\
0   , \qquad r  > r_{c1} \;,
\end{array} \right.
\end{equation}

\noindent where $r_{c1} = 2.0$ and $\varepsilon_{1} 
= A_1\left( \sigma/r_{c1} \right)^{6} 
+ A_2(r_{c1}/\sigma)$. For the R24 potential we have

\begin{equation}\label{eq_potentialr24}
U_{R24}^* = \frac{U_{R24}}{\epsilon} = \left\{ \begin{array}{ll}
  B_1\left( \sigma/r \right)^{24} - \varepsilon_{2}  , \qquad r \le r_{c2} \\
0   , \qquad r  > r_{c2} \;,
\end{array} \right.
\end{equation}

\noindent where $r_{c2} = 1.50$ and 
$\varepsilon_{2} = B_1\left( \sigma/r_{c2} \right)^{24}$.  
The Weeks-Chandler-Andersen 
Lennard-Jones potential (WCA) is given by
\begin{equation}\label{eq_potential2}
U_{WCA}^* = \frac{U_{WCA}}{\epsilon}= \left\{ \begin{array}{ll}
 U_{{\rm {LJ}}}(r) - U_{{\rm{LJ}}}(r_{ c3})  , \qquad r \le r_{ c3} \\
0   , \qquad r  > r_{ c3} \;,
\end{array} \right.
\end{equation}

\noindent where $U_{LJ}(r)$ is a standard 12-6 LJ potential and 
$r_{c3} = 2^{1/6}\sigma$. 

Two hydrophilic potentials are analyzed: one weakly hydrophilic, WAT, and another more 
strong, SAT. The hydrophilic WAT potential is given by

\begin{equation}\label{eq_potentialWAT}
U_{WAT}^* = \frac{U_{WAT}}{\epsilon} = \left\{ \begin{array}{ll}
 C_1\left[ (\sigma/r)^{12}-(\sigma/r)^{6}\right] + 
C_2\left( r/\sigma\right) - \varepsilon_{4}  , \qquad r \le r_{c4} \\
0   , \qquad r  > r_{c4} \;,
\end{array} \right.
\end{equation}

\noindent where $r_{c4} = 1.5$ and $\varepsilon_{4} 
= C_1\left[ (\sigma/r_{c4})^{12}-(\sigma/r_{c4})^{6}\right] 
+ C_2(r_{c4}/\sigma)$. 

The equation for the SAT potential is

\begin{equation}\label{eq_potentialSAT}
U_{SAT}^* = \frac{U_{SAT}}{\epsilon} = \left\{ \begin{array}{ll}
D_1\left[(\sigma/r)^{12}-(\sigma/r)^{6}\right]
 + D_2\left( r/\sigma\right) - \varepsilon_{5}  , \qquad r \le r_{c5} \\
0   , \qquad r  > r_{c5} \;,
\end{array} \right.
\end{equation}

\noindent where $r_{c5} = 2.0$ and $\varepsilon_{5} = 
D_1\left[(\sigma/r_{c5})^{12}-(\sigma/r_{c5})^{6}\right] + 
D_2(r_{c5}/\sigma)$.

The parameters are illustrated in table \ref{table1}.
 \begin{table} [!htb]
  \caption{Parameters of the particle-plate potentials.} 
\vspace{0.5cm}
  \begin{tabular}{c|c|c}
  \hline\hline
\ \ Potential \ & \ \ Parameters values \ & \ \ Parameters values  \ \ \tabularnewline \hline
\ \ R6 \ & \ \ $A_1 = 4.0$       \ & \ \    $A_2 = 0.1875$       \ \ \tabularnewline
\ \ R24 \ & \ \ $B_1 = 4.0$       \ & \ \    $--$       \ \ \tabularnewline
\ \ WAT \ & \ \ $C_1 = 1.0$       \ & \ \    $C_2 = 0.289$       \ \ \tabularnewline 
\ \ SAT \ & \ \ $D_1 = 1.2$       \ & \ \    $D_2 = 0.0545$       \ \ \tabularnewline \hline\hline

  \end{tabular}\label{table1}
 \end{table}

\begin{figure}[!htb]
  \begin{centering}
\includegraphics[clip=true,width=10cm]{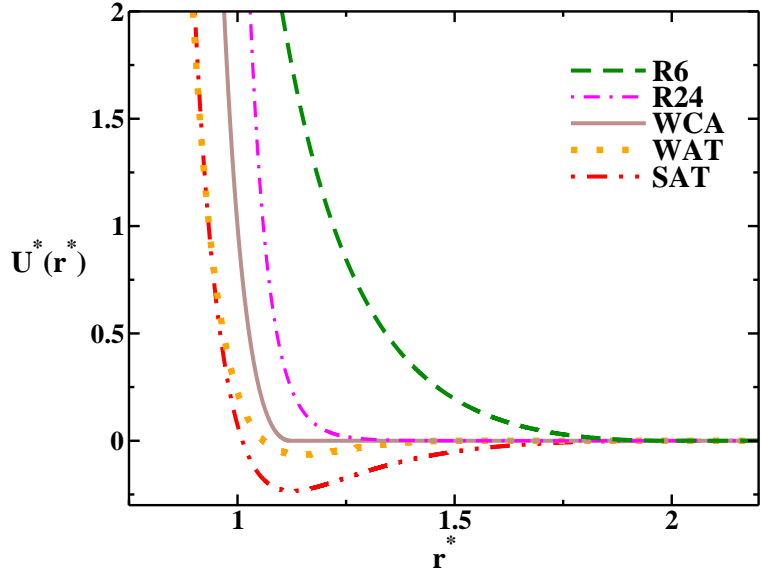}\par
    \par
  \end{centering}
  \caption{Particle-plate interaction potentials: the three purely
repulsive, R6(dashed line), R24(dotted-dashed line), and
WCA(solid line), and the attractive, WAT(dotted line) and SAT(double-dotted-dashed line).}
  \label{potential_plates}
\end{figure}

The figure \ref{potential_plates} illustrates the particle-plate interaction 
potentials.

\section{\label{sec:simulation} The Methods and Simulation Details}

The systems are formed by $N$ particles confined in $z$ direction by two rough plates 
with area $L^2$,
located each one at $z = 0$ and $z = d$. The position of each plate is fixed. To 
simulate infinite systems in $x$ and $y$ directions, in order
to have the thermodynamic limit, we employed periodic boundary conditions in them. 
Due to the empty space near to each plates, the distance $d$ between them needs to be 
corrected to an effective distance \cite{Ku05, Ku07}, $d_e$, that can be approach by 
$d_e \approx d -\sigma$. Consequently, the effective density will be $\rho_e = N/(d_eL^2)$.

We use molecular dynamic simulations at the NVT-constant ensemble to study the 
problems suggested.
To keep fixed the temperature, we used the 
Nose-Hoover~\cite{nose_hoover_85,nose_hoover_86} thermostat,
with coupling parameter $Q = 2$. The particle-particle interaction was done 
until the cutoff radius $r_c = 3.5$
and the potential was shifted in order to have $U = 0$ at $r_c$. 

For all the particle-plate interaction potentials, we study systems with plates 
separated by distances 
$d^* = d/\sigma = 4.2$, $6.0$ and $10.0$. The properties of each case were studied 
for several temperatures 
and densities to obtain the full phase diagrams. We use $N = 507$ particles for 
systems at $d^* = 4.2$ and $d^* = 6.0$, 
and $N = 546$ particles for systems at $d^* = 10.0$. The initial configuration 
was set on solid structure and the equilibrium states 
reached after $2 \times 10^6$ steps, followed by $4 \times 10^6$ 
simulation run. The time step was $0.001$ in reduced units and the 
average of the physical quantities were get with $50$ descorrelated 
samples. We used the behavior of the energy after the equilibrium states
and the parallel and perpendicular pressure as function of density 
to check the thermodynamic stability.

This kind of system requires the division of the thermodynamic averages in the components 
parallel and perpendicular to the plates. In systems with this geometry, the Helmholtz 
free energy is given in terms of area, $A=L_xL_y$, and distance
between the plates, $L_z$ \cite{Me99}. Considering the periodic boundary conditions 
in the plane, 
the system is extensive just in the area but not in the distance between the plates. 
Therefore, only the parallel pressure can be regarded as a thermodynamic
quantity and it might scale as the experimental pressure. Considering that, we
are interested just in the quantities related to parallel direction.

The parallel pressure was calculated using the Virial expression for the $x$ and $y$ 
directions~\cite{Ku05,Ha08,Me99,Gi09,Ku09,Ku07}. The dynamic of the systems was 
studied by lateral diffusion
coefficient, $D_{\parallel}$, related with the mean square 
displacement (MSD) from Einstein relation,
\begin{eqnarray}
\label{difusao_lateral}
D_{\parallel} = \lim_{\tau\to\infty} 
\frac{\langle\Delta r_{\parallel}(\tau)^2\rangle}{4 \tau},
\end{eqnarray}
\noindent where $r_{\parallel} = (x^2+y^2)^{1/2}$ is the distance 
between the particles parallel to the plates. 

We also studied the structure of the systems by lateral radial 
distribution function, $g_{\parallel}(r_{\parallel})$. We calculate the 
$g_{\parallel}(r_{\parallel})$ in specific regions between the 
plates. An usual definition for $g_{\parallel}(r_{\parallel})$ is 
\begin{eqnarray}
\label{gr_lateral}
g_{\parallel}(r_{\parallel}) \equiv \frac{1}{\rho ^2V}
\sum_{i\neq j} \delta (r-r_{ij})\left [ \theta\left( \left|z_i-z_j\right| 
\right) - \theta\left(\left|z_i-z_j\right|-\delta z\right) \right].
\end{eqnarray}
The $\theta(x)$ is the Heaviside function and it restricts the sum of 
particle pairs in the same slab of thickness $\delta z = 1$. We need to 
compute the number of particles for each region and the normalization 
volume will be cylindrical. The $g_{\parallel}(r_{\parallel})$ is 
proportional to the probability of finding a particle at a distance 
$r_{\parallel}$ from a referent particle.

All physical quantities are shown in reduced units~\cite{Al87} as

\begin{eqnarray}
d^* &=& \frac{d}{\sigma} \nonumber \\
\tau^* &=& \frac{(\epsilon/m)^{1/2}}{\sigma} \tau \nonumber \\
T^* &=& \frac{k_B}{\epsilon}T \nonumber \\
P_{\parallel, \perp}^* &=& \frac{ \sigma^3 }{\epsilon}P_{\parallel, \perp} \nonumber \\
\rho^* &=& \sigma ^3\rho \nonumber \\
D_{\parallel}^* &=& \frac{(m/\epsilon)^{1/2}}{\sigma}D_{\parallel} \;\; .
\end{eqnarray}

\section{\label{sec:results} Results}

\subsection*{The Structure}

First, we check the effects of decreasing the plates distance
and the hydrophobicity in the number of layers of water and its structure.
For that purpose we focus in three layer distances 
$d^* = 10.0, 6.0$ and $4.2$ in which we observe the 
presence of 
five, three and two layers respectively  for all the particle-plate potentials.

\begin{figure}[!htb]
 \centering

 \begin{tabular}{ccc}
 \includegraphics[clip=true,width=8cm]{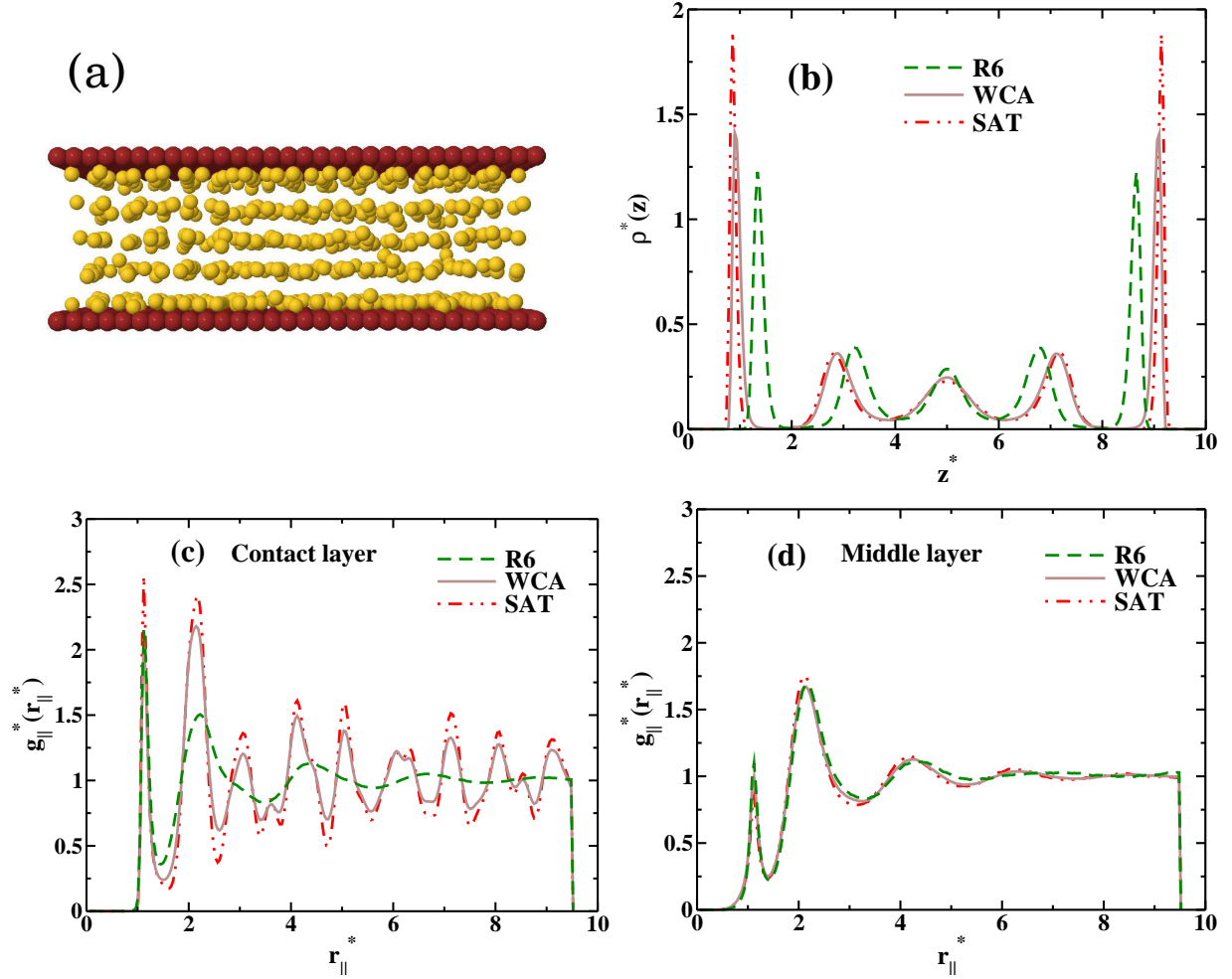} 
 \includegraphics[clip=true,width=8cm]{hist_lz10_140_21_0.eps} \\
 \includegraphics[clip=true,width=8cm]{comp_gr_lz10_wall_140_20_0.eps} 
 \includegraphics[clip=true,width=8cm]{comp_gr_lz10_140_20_0.eps} 
\tabularnewline
 \end{tabular}\par
 \caption{Systems with plates separated by a distance $d^* = 10.0$ 
at $\rho^* = 0.168$ and $T^* = 0.140$. (a) Snapshot showing the five layers for the
WCA case. (b)Transversal density versus $z$ for systems confined
by the R6, WCA and SAT potentials. Radial distribution function versus distance for
the (c) contact and (d) middle layers. The confinements by the R24 and WAT
potentials have similar results than the WCA and are not shown for simplicity.}
\label{gr_compare10}
\end{figure}

The figure~\ref{gr_compare10} illustrates the structure for the distance 
between the plates 
$d^* = 10.0$ at $\rho^* = 0.168$ and $T^* = 0.140$ only for the three
cases R6, WCA and SAT for simplicity. 
In the figure~\ref{gr_compare10}(a) the snapshot shows the 
structure with  five layers 
(only the WCA for simplicity). In the figure~\ref{gr_compare10}(b) the 
density at the $z$ direction is plotted against $z$, showing that
for the attractive potential the contact layer is 
closer to the plates when compared with the purely repulsive
particle-plate potentials. The distance
between the layers is arranged to minimize
the particle-particle interaction illustrated in 
the figure~\ref{potential_alan}, while the distance between
the plate and the contact layer to minimize
the particle-plate interaction. This simple geometrical
arrangement is robust for all the potentials and
as we shall see below for various plate-plate distances.
The figures~\ref{gr_compare10}(c) and (d)
illustrates the radial distribution functions for the
contact and middle  layers respectively. While the 
contact layers show the presence of an amorphous-like structure,
the middle layers are liquid. The only case in which both layers are 
liquids is the repulsive case R6. This result is in agreement with SPC/E simulation of
Gallo et al.~\cite{Ga02b}.

\begin{figure}[!htb]
 \centering
 \begin{tabular}{ccc}
 \includegraphics[clip=true,width=8cm]{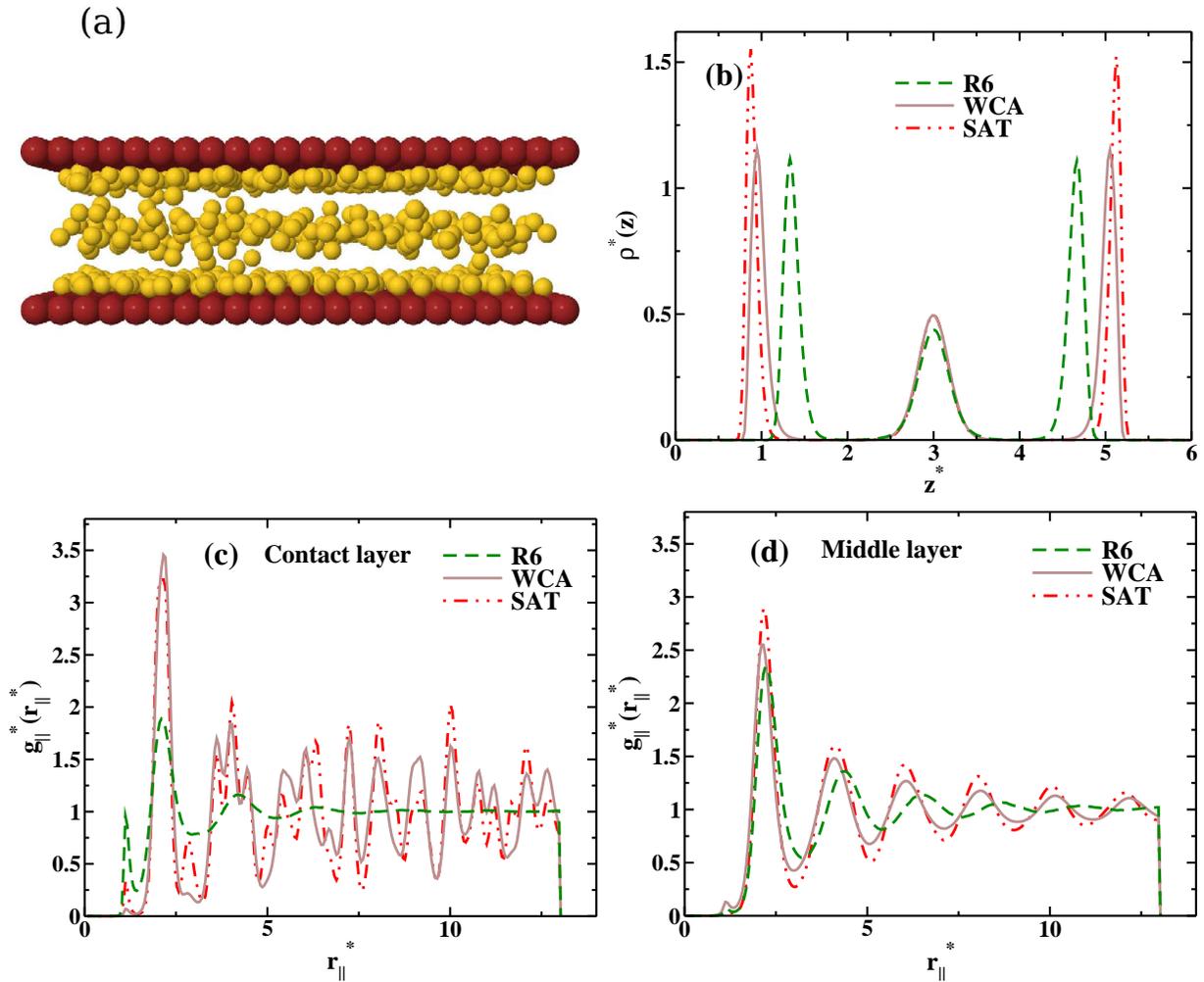} 
 \includegraphics[clip=true,width=8cm]{hist_lz6_140_28_0.eps}\\
 \includegraphics[clip=true,width=8cm]{comp_gr_lz6_wall_140_27_0.eps} 
 \includegraphics[clip=true,width=8cm]{comp_gr_lz6_140_27_0.eps} 
\tabularnewline
 \end{tabular}\par
 \caption{Systems with plates separated by a distance $d^* = 6.0$ 
at $\rho^* = 0.150$ and $T^* = 0.140$. (a) Snapshot showing the three layers for the
WCA case. (b)Transversal density versus $z$ for systems confined
by the R6, WCA and SAT potentials. (c) Radial distribution function versus distance for
the contact and (d) middle layers. The confinements by the R24 and WAT
potentials have similar results than the WCA and are not shown for simplicity.}\label{gr_compare6}
\end{figure}

The figure~\ref{gr_compare6}  illustrates 
the system for plates separated by a 
distance $d^* = 6.0$ at $\rho^* = 0.150$ and $T^* = 0.140$. In the
figure~\ref{gr_compare6}(a) the snapshot shows
the structure with three layers (only the WCA for simplicity), two
contact layers and one middle layer. 
In the figure~\ref{gr_compare6}(b) the density 
versus $z$ indicates that as the plates becomes more 
attractive, particles are pushed toward them. The 
figures~\ref{gr_compare6}(c) and (d)
show that for $d^* = 6.0$ the contact layer presents
an amorphous-like structure while  
the middle layer is fluid. This observation is true for all the 
particle-plate potentials with the
exception of the R6 which is fluid at the contact
and middle layers. 

\begin{figure}[!htb]
 \centering
 \begin{tabular}{cc}
 \includegraphics[clip=true,width=8cm]{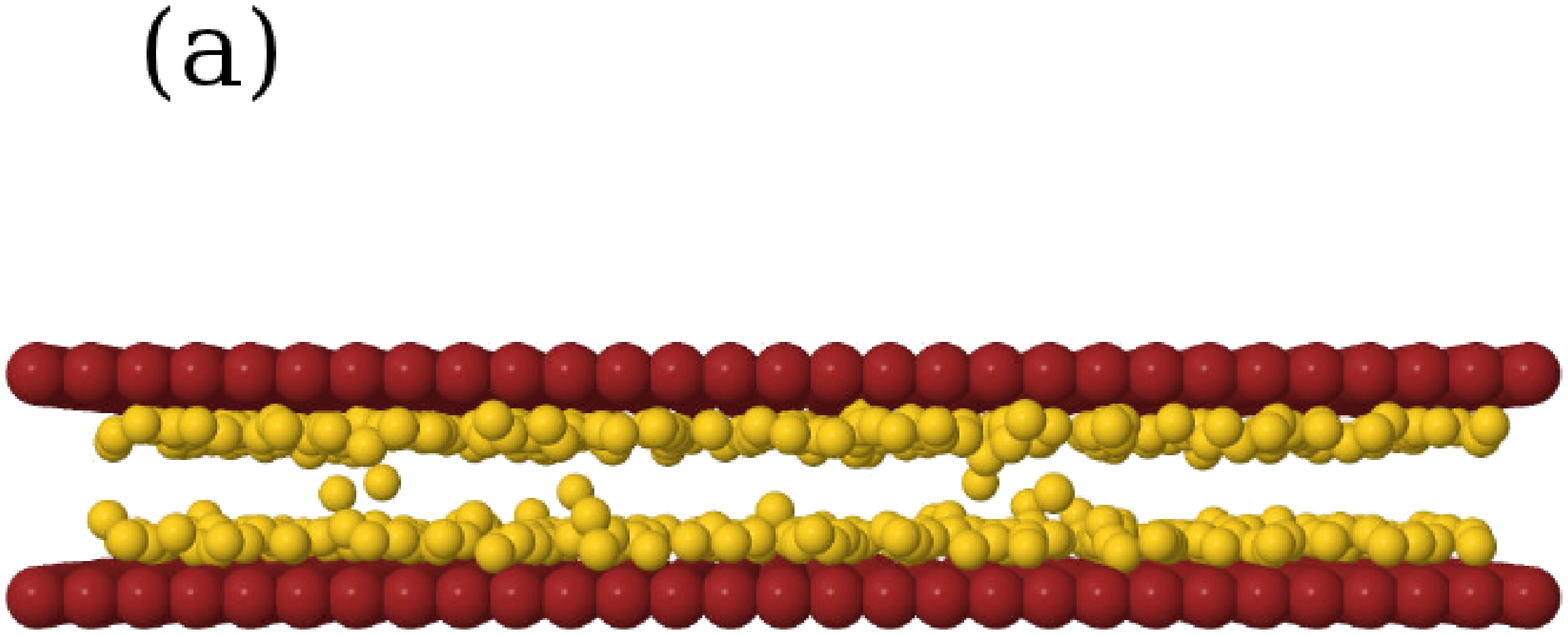} 
 \includegraphics[clip=true,width=8cm]{hist_comp_140_32_0.eps}\\
 \includegraphics[clip=true,width=8cm]{comp_gr_lz4_140_32_0.eps}
 \includegraphics[clip=true,width=8cm]{comp_gr_lz4_250_32_0.eps} 
\tabularnewline
 \end{tabular}\par
 \caption{Systems with plates separated by a distance $d^* = 4.2$ 
at $\rho^* = 0.165$. (a) Snapshot 
showing the two layers for the
WCA case at $T^*=0.140$. (b)Transversal density versus $z$ for systems confined
by the R6, WCA and SAT potentials. Radial distribution function versus distance for
one contact layer for (c) $T^*=0.140$ and for (d)  $T^*=0.250$. The confinements by the R24 and WAT
potentials have similar results than the WCA and are not shown for simplicity.}\label{gr_compare4_2}
\end{figure}

For the distance $d^* = 4.2$ and  $\rho^* = 0.165$ two temperatures
were analyzed, $T^* = 0.140$ and $T^* = 0.250$. The
figure~\ref{gr_compare4_2}(a) shows a snapshot of the system (only
for the WCA for simplicity) indicating the presence
of two layers. The figure~\ref{gr_compare4_2}(b) shows the density at 
the $z$ direction. Similarly to  the $d^*=10$ and $6.0$ cases the 
mainly effect of hydrophobicity is to have the two layers closer
to the wall than in the case of the hydrophilic wall. The figure~\ref{gr_compare4_2}(c) and (d)
shows the radial distribution function 
for two temperatures  $T^* = 0.140$ and 
$T^* = 0.250$ respectively. While $T^* = 0.140$ shows an amorphous-like
structure for the WCA and SAT potentials, $T^* = 0.250$ is liquid, indicating that the system
melts at an intermediate temperature. The R6 potentials exhibits a liquid-like behavior for
both temperatures.

\begin{figure}[!htb]
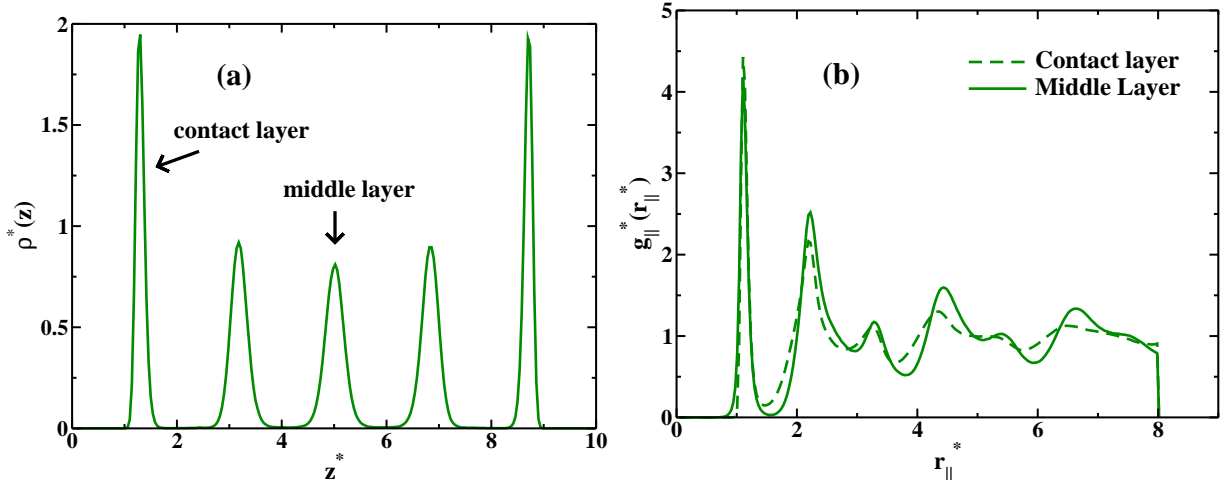

\centering
\begin{tabular}{cc}
\includegraphics[clip=true,width=8cm]{hist_lz10_18_0_140.eps} &
\includegraphics[clip=true,width=8cm]{comp_gr_lz10_18_0_140_layers.eps} 
\tabularnewline
\end{tabular}\par
\caption{Transversal density versus $z$ in (a) and $g_{||}(r_{||})$ in (b) for $d^* = 10.0$, 
$\rho^* = 0.209$ and $T^* = 0.140$. The middle layer is more structured 
than the contact layer, which is in agreement with some experimental results for
hydrophobic confinement.}\label{gr_mc}
\end{figure}

In all the cases showed above, the purely repulsive potential R6 has no crystalline layer.
This suggests that the crystallization in this case occur at higher pressures. In order 
to check that, the case at $d^* = 10.0$ is analyzed for $\rho^* = 0.210$ in comparison
with $\rho^* = 0.168$, shown in figure~\ref{gr_compare10}. In the figure~\ref{gr_mc} (a) 
the transversal density versus $z$ is shown for a system with plates separated
by $d^* = 10.0$ at $\rho^* = 0.210$ and $T^* = 0.140$, showing the five layers
formed, and in (b), we have the 
$g_{||}(r_{||})$ of the contact and the middle layers. Both layers represent 
amorphous states and, besides that, it is possible to see that the middle layer is smoothly more structured 
than the contact layer.
This is an interesting and an important observation because some experimental 
results say that the 
middle layer crystallizes before than the contact layer \cite{Er11, Mo97}. This 
result is obtained just
for the R6 potential, which lead us again to the conclusion that this potential 
is the best to reproduce 
the structure related in some experiments for the hydrophobicity.

\begin{figure}[!htb]
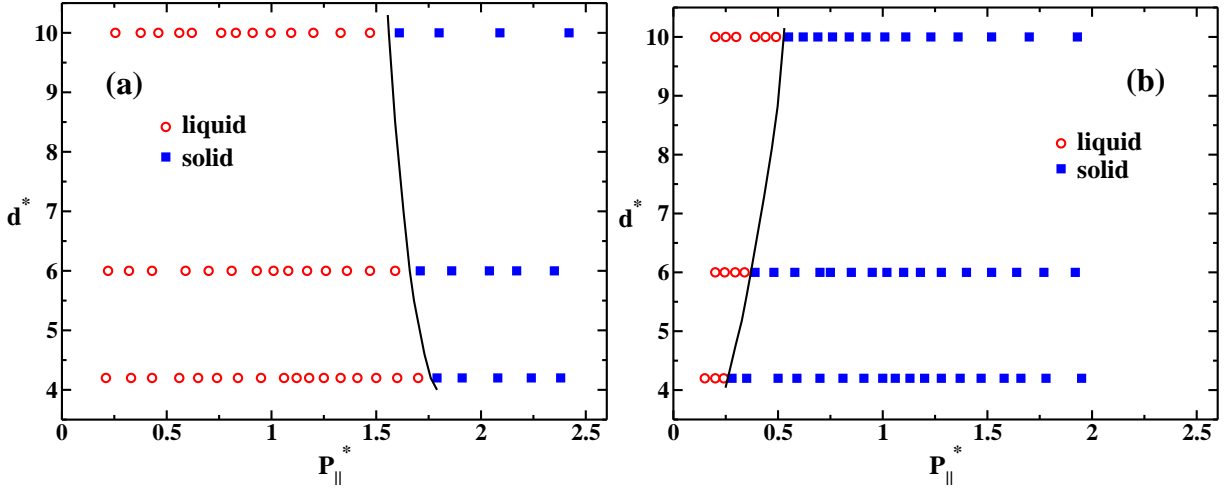

 \centering
 \begin{tabular}{ccc}
 \includegraphics[clip=true,width=8cm]{d_P_r6140.eps}
 \includegraphics[clip=true,width=8cm]{d_P_atm140.eps}
\tabularnewline
 \end{tabular}\par
 \caption{Distance $d^*$ between the plates as function of density of the system
 for (a) R6 and (b) SAT confinements at $T^* = 0.140$. The open circles indicate liquid-state
points while filled squares indicate solid-state points. The black lines are an approximate boundary between
the liquid and solid-states.}\label{melt1}
\end{figure}

Our results, comparing the different potentials and plates distances,
indicate that the hydrophobicity has little effect in number of layers
that is defined by the distance $d^*$ between the plates.
The crystallization of the contact layer, however, seems to 
be dependent both of the particle-plate interaction and of the
distance between the plates. 
In order to explore in detail the process of crystallization
of the contact layer we analyze the phase behavior of the confined systems 
for the R6 and SAT potentials. The figure~\ref{melt1} shows 
the phase behavior of the systems confined by the (a) R6 and 
(b) SAT potentials at $T^* = 0.140$. The open circles indicate liquid-state
points while filled squares indicate solid-state points. An approximate 
boundary between the liquid and solid-states are indicated by the black lines in the figures.
For this specific temperature, our results suggest that the melting pressure
decreases with $d^*$ for the hydrophobic potentials and increases with $d^*$ 
for the hydrophilic potentials. This result is consistent with the liquid-gas observations
in the SPC/E confined model \cite{Gi09a}.

\begin{figure}[!htb]
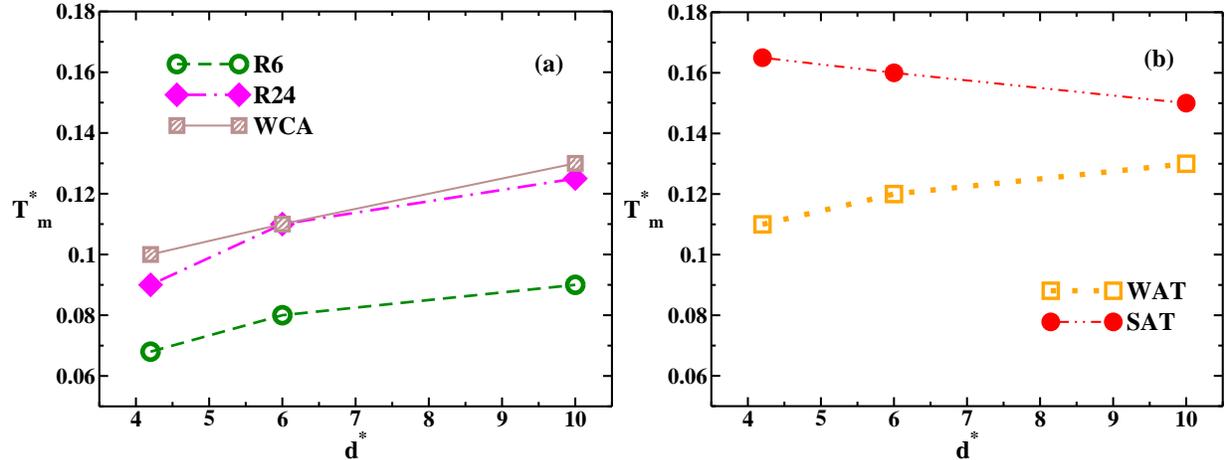

 \centering
 \begin{tabular}{ccc}
 \includegraphics[clip=true,width=8cm]{melting_rep.eps}
 \includegraphics[clip=true,width=8cm]{melting_act.eps} 
\tabularnewline
 \end{tabular}\par
 \caption{Melting temperature as function of 
separation $d^*$ between the plates for (a) repulsive (R6, R24 and WCA) and
(b) attractive (WAT and SAT) potentials. Systems at $\rho^* = 0.176$.}\label{melt2}
\end{figure}

The figure~\ref{melt2} shows the melting temperature of the
systems at $\rho^* = 0.176$ for (a) repulsive potentials (R6, R24 and WCA)
and for (b) attractive potentials (WAT and SAT).
For temperatures $T^* > T^*_m$, all the system 
are in liquid-state, while for $T^* < T^*_m$ a crystallization occurs at least 
for the contact layers. The SAT potential crystallizes more easily than the other cases
and has a peculiar behavior with the distance $d^*$ between the plates.
The crystallization for the SAT potentials occurs more easily as the degree 
of confinement increases (decreasing of $d^*$), while other cases show 
the opposite behavior. Our results are in agreement with simulations and experiments for 
water confined in hydrophobic and hydrophilic nanopores \cite{Mo12, De10}.

\subsection*{\label{thermo} Diffusion and density anomalous behavior}

\begin{figure}[!htb]
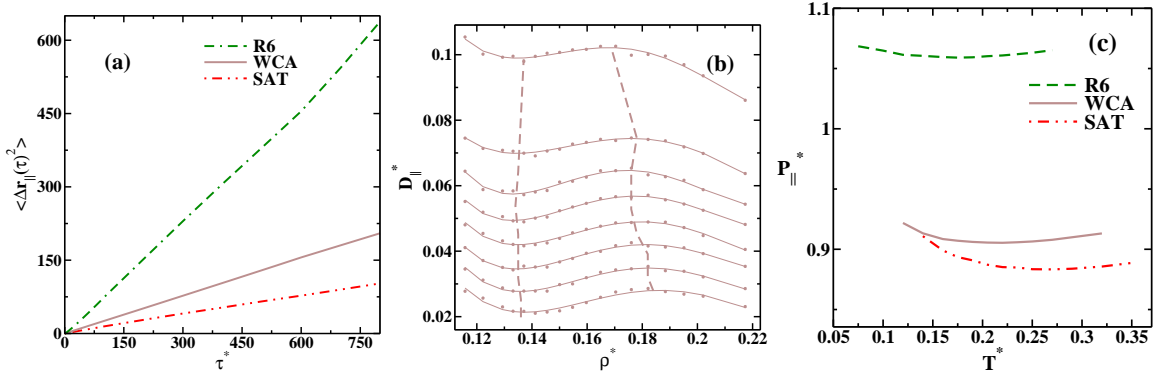

 \centering
 \begin{tabular}{ccc}
 \includegraphics[clip=true,width=5cm]{r2_comp_32_0_250_4_2_2.eps} 
 \includegraphics[clip=true,width=5cm]{diff_4_2_wca.eps} 
 \includegraphics[clip=true,width=5cm]{comp_isochore_4_2_32_0.eps} 
\tabularnewline
 \end{tabular}\par
 \caption{Systems for
$d^* = 4.2$, $\rho^* = 0.165$ and $T^* = 0.250$. (a) Mean
square displacement versus time for R6, WCA and SAT potentials.
(b) Diffusion coefficient  versus density for the 
WCA potential at fixed temperatures 
$T^* = 0.175$, $0.190$, $0.205$, $0.220$, $0.235$, $0.250$, $0.270$ 
and $0.320$ from the bottom to the top. (c) Isochore $\rho^* = 0.165$ at
the pressure-temperature phase diagram for R6, WCA and SAT potentials.
The R24 and WAT potentials are intermediate cases and are not shown for simplicity.
} \label{r2_comp}
\end{figure}

In this subsection we analyze the effect of hydrophobicity and changing the
plates distances in the location in the pressure-temperature phase
diagram of the diffusion and density anomalies. 
The Figure~\ref{r2_comp} (a) shows a comparison between the mean square 
displacement parallel to the plates
at $d^* = 4.2$, $\rho^* = 0.165$ and $T^* = 0.250$. The plot
shows that the  mobility is higher
for hydrophobic than hydrophilic particle-plate interactions. This 
result is consistent with the layer density
illustrated in the Figure~\ref{gr_compare4_2}(b) that shows that attractive particle-plate
interactions leave more space for the layers.  In the Figure~\ref{r2_comp}(b)
the diffusion coefficient is shown as a function of the density for 
$d^* = 4.2$ and WCA confining potential, illustrating
the presence of a region where diffusion increases with the increase
of the density what is defined as diffusion anomalous region (region
between the dashed lines).
This anomalous behavior is also present for the other distances $d=6.0$ and 
$10.0$ and other particle-plate potentials. 

\begin{figure}[!htb]
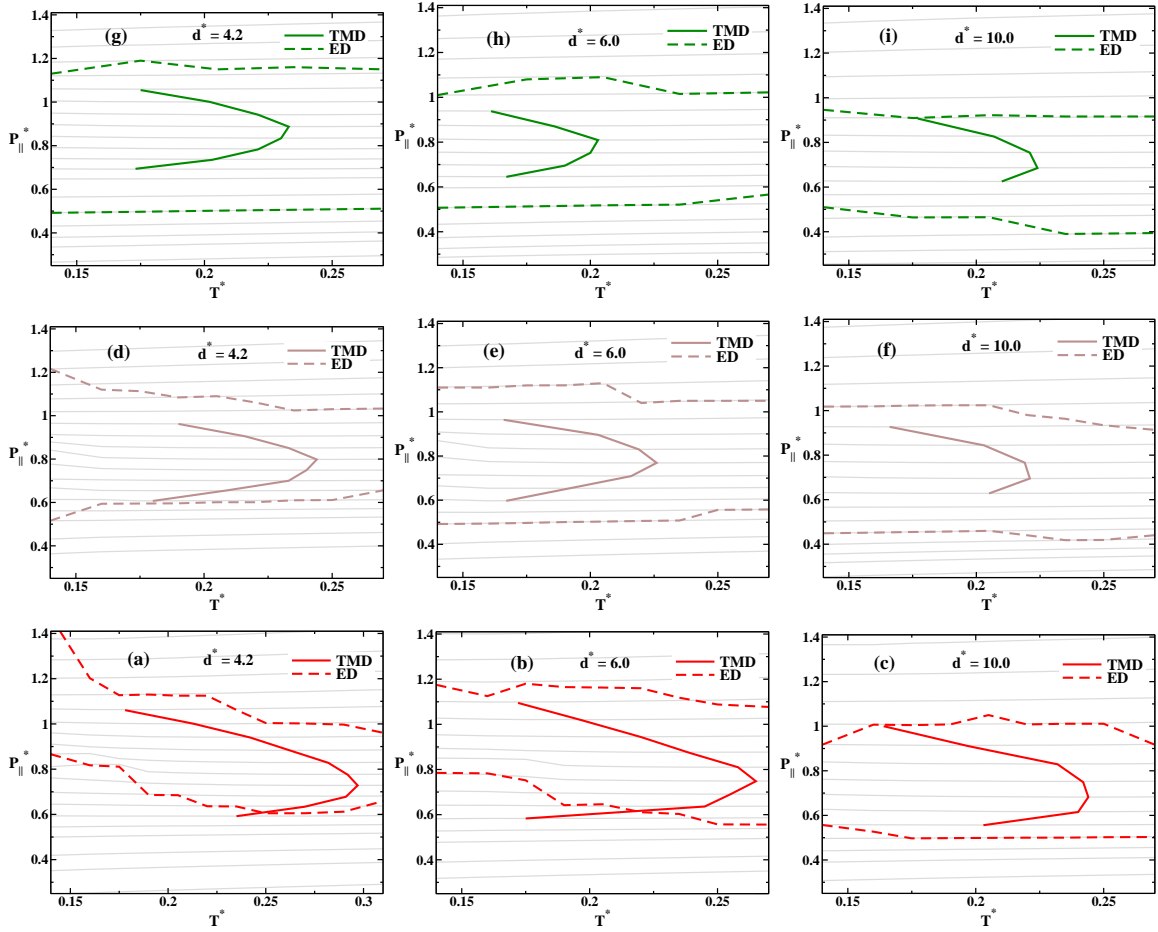

 \centering
 \begin{tabular}{ccc}
 \includegraphics[clip=true,width=5cm]{phase_diagram_lz4_2_r6.eps} 
 \includegraphics[clip=true,width=5cm]{phase_diagram_lz6_r6.eps} 
 \includegraphics[clip=true,width=5cm]{phase_diagram_lz10_r6.eps} \\
 \includegraphics[clip=true,width=5cm]{diagram_p_parallel_lz4_2_completo_PB.eps} 
 \includegraphics[clip=true,width=5cm]{diagram_p_parallel_lz6_completo.eps} 
 \includegraphics[clip=true,width=5cm]{diagram_p_parallel_lz10_completo_PB.eps} \\
 \includegraphics[clip=true,width=5cm]{phase_diagram_lz_4.2_atm.eps} 
 \includegraphics[clip=true,width=5cm]{phase_diagram_lz_6_atm.eps} 
 \includegraphics[clip=true,width=5cm]{phase_diagram_lz10_atm.eps}
\tabularnewline
 \end{tabular}\par
 \caption{Parallel pressure versus temperature phase diagrams for (a)-(c), R6 (d)-(f) WCA 
and (g)-(h) SAT potentials. For all the plots the solid line is the 
TMD line and the dashed line is the extremum diffusion coefficient line. 
The range of densities are $0.089 \le \rho^* \le 0.182$ for systems at $d^* = 4.2$, 
$0.087 \le \rho^* \le 0.176$ for $d^* = 6.0$
and $0.083 \le \rho^* \le 0.168$ for $d^* = 10.0$.}\label{phase_diagrams}
\end{figure}

Next, we test our system 
for the presence of  the temperature of maximum density (TMD).
The TMD lines can be found computing 
$(\partial P_{||}/\partial T)_{\rho} = 0$, corresponding 
to minimum of the isochores. A comparison between the same
isochore ($\rho^* = 0.165$) for each potential is given in the 
Figure~\ref{r2_comp} (c) for $d^* = 4.2$. The temperature of maximum 
density  decreases and its pressure increases
as the system becomes more hydrophobic.
The pressure increase can be understood in terms of the decrease
of effective volume for hydrophobic plates as shown in figure~\ref{gr_compare4_2}(b)
with corresponding increase of pressure. The decrease of the 
TMD with hydrophobicity can be understood as follows. In our 
effective model the two length scales represent the bond and 
non-bonding cluster of molecules. As temperature increases the 
number of clusters with ``non-bonding molecules grow
while the number of clusters with ``bonding'' molecules decreases. The
TMD is the temperature in which the two distributions become equivalent.
In the confined system the wall repulsion favor the ``non-bonding length
scale and the TMD happens at lower values. The Figure~\ref{phase_diagrams}
shows the parallel pressure versus temperature phase diagram 
for (a)-(c) R6, (d)-(f) WCA and (g)-(h) SAT potentials. The dashed
lines comprises the diffusion anomaly and the solid lines indicate the
density anomaly for each case. For all the cases studied, the hierarchy
of the anomalies are observed.

\begin{figure}[!htb]
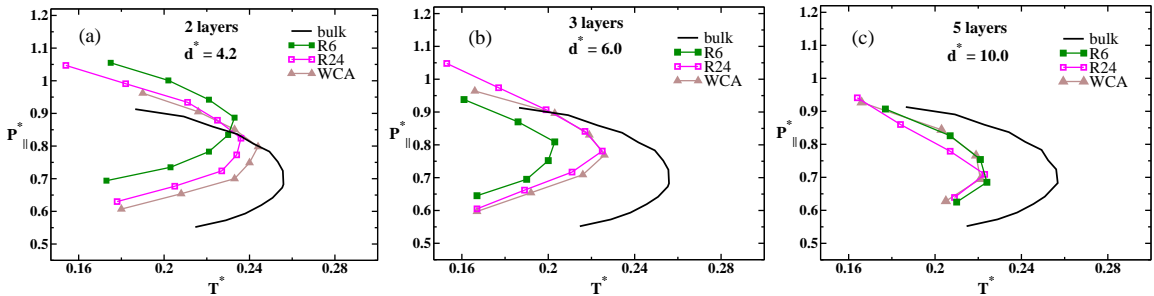

 \centering
 \begin{tabular}{ccc}
 \includegraphics[clip=true,width=5cm]{tmds_lz4_2_hb.eps} &
 \includegraphics[clip=true,width=5cm]{tmds_lz6_hb.eps} &
 \includegraphics[clip=true,width=5cm]{tmds_lz10_hb.eps} 
\tabularnewline
 \end{tabular}\par
 \caption{Pressure versus temperature phase diagram illustrating
the TMD line for hydrophobic confinement, R6, R24 and WCA potentials, 
for (a) $d^* = 4.2$, (b) $d^* = 6.0$ and  (c) $d^* = 10.0$.}\label{tmds_hb}
\end{figure}

Confirming the 
scenario we describing above, Figure~\ref{tmds_hb} illustrates
the  TMD lines for the hydrophobic particle-plate interaction potentials 
for
(a) $d^* = 4.2$, (b) $d^* = 6.0$ and (c) $d^* = 10.0$. The TMD lines 
are shifted to lower temperatures in relation to bulk system 
as the distance between the plates is decreased. This result is
consistent with atomistic models~\cite{Ku05,Gi09}.

\begin{figure}[!htb]
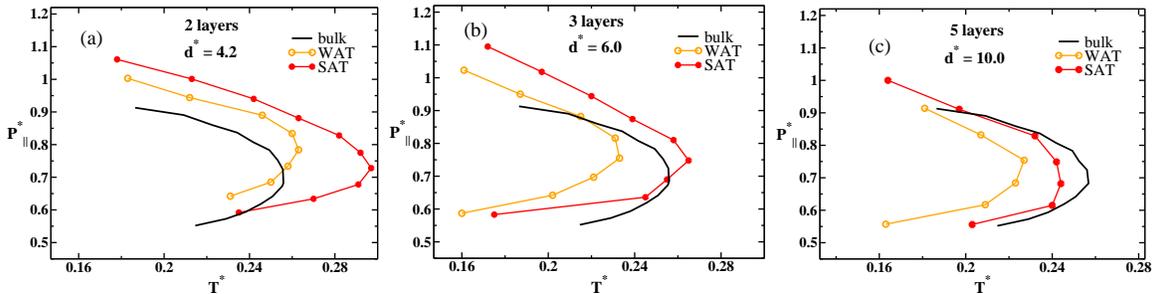

 \centering
 \begin{tabular}{ccc}
 \includegraphics[clip=true,width=5cm]{tmds_lz4_2_hl.eps} &
 \includegraphics[clip=true,width=5cm]{tmds_lz6_hl.eps} &
 \includegraphics[clip=true,width=5cm]{tmds_lz10_hl.eps} 
\tabularnewline
 \end{tabular}\par
 \caption{Pressure versus temperature phase diagram 
illustrating the TMD line for hydrophilic confinement, WAT and SAT potentials, for (a) $d^* = 4.2$, (b) $d^* = 6.0$ and  (c) $d^* = 10.0$.}\label{tmds_hl}
\end{figure}

Figure~\ref{tmds_hl} shows the TMD lines for the hydrophilic particle-plate
interaction potentials for different
plates separations. For these cases
the TMD moves to higher temperatures when compared
to the bulk values as the distance between the plates is
decreased. This result is consistent
with atomistic models~\cite{Ca09}.

\section{\label{sec:conclusions} Conclusions}

In this paper we have explored the effect of the confinement in the thermodynamic, dynamic
and structural properties of a core-softened potential designed to reproduce the anomalies present
in water. 

We have shown that both hydrophobic and hydrophilic walls change the melting, the TMD and the extrema diffusivity
temperatures. While melting is suppressed by hydrophobic walls, crystallization happens for hydrophilic confinement
at higher temperatures if the walls would be attractive enough. 

Our results suggest that layering, crystallization and thermodynamic and dynamic anomalies are governed by the competition
between the two length scales that characterize our model and the particle-plate interaction length. These
results are consistent with atomistic models \cite{Gi09a, Mo12, So11, Ku05}, however due the simplicity of the simulation we were able to 
explore a large variety of potentials to confirm our assumption that a simple competition between scales not only
is able to reproduce the water anomalies but to capture the confinement phase diagram.
\section*{ACKNOWLEDGMENTS}

We thank for financial support the Brazilian science agencies, CNPq 
and Capes. This work is partially supported by CNPq, INCT-FCx. We also 
thank to CEFIC - Centro de F\'isica Computacional of Physics Institute 
at UFRGS, for the computer clusters.

\vspace{1cm}
\bibliographystyle{aip}

\end{document}